\documentclass[useAMS,usenatbib,letters]{mn2e}
\usepackage{epsfig}
\usepackage{epstopdf}
\usepackage{graphicx,amssymb,supertabular,longtable,lscape}
\usepackage{amsmath}
\usepackage{textcomp}
\usepackage{colortbl, xcolor}
\usepackage{multirow}
\usepackage{threeparttable}
\usepackage{color}
\definecolor{mycol}{cmyk}{0.00, 0.25, 0.00, 0.00}
\def\apj{\mbox{ApJ}}
\def\apjl{\mbox{ApJL}}
\def\apjs{\mbox{ApJS}}
\def\aass{\mbox{A\&AS}}
\def\mnras{\mbox{MNRAS}}
\def\aj{\mbox{AJ}}
\def\araa{\mbox{ARA\&A}}

\def\nat{\mbox{Nature}}
\def\aap{\mbox{A\&A}}

\title[Disc scalelengths out to redshift 5.8]
{Disc scalelengths out to redshift 5.8}

\author[Fathi et al.]{Kambiz Fathi$^{1,2,}$\thanks{e-mail: kambiz@astro.su.se}, Michael Gatchell$^{1,3}$, Evanthia Hatziminaoglou$^{4}$, Benoit Epinat$^{5,6}$\\ \ \\
$^{1}$Stockholm Observatory, Department of Astronomy, Stockholm University, AlbaNova Centre, 106 91 Stockholm, Sweden\\
$^{2}$Oskar Klein Centre for Cosmoparticle Physics, Stockholm University, 106 91 Stockholm, Sweden\\
$^{3}$Department of Physics, Stockholm University, AlbaNova Centre, 106 91 Stockholm, Sweden\\
$^{4}$ESO, Karl-Schwarzschild-Str. 2, 85748 Garching bei M\"unchen, Germany\\
$^{5}$Institut de Recherche en Astrophysique et Planetologie, Universite de Toulouse, CNRS,
14 avenue Edouard Belin, 31400 Toulouse, France\\
$^{6}$Laboratoire dÕAstrophysique de Marseille, Universite de Provence, CNRS, 38 rue Frederic
Joliot-Curie, F-13388 Marseille Cedex 13, France}
\begin{document}

\pagerange{\pageref{firstpage}--\pageref{lastpage}} \pubyear{2010}

\maketitle

\label{firstpage}

\begin{abstract}
We compute the exponential disc scalelength for 686 disc galaxies with spectroscopic redshifts out to redshift 5.8 based on Hubble Space Telescope archival data. We compare the results with our previous measurements based on 30000 nearby galaxies from the Sloan Digital Sky Survey. Our results confirm the presence of a dominating exponential component in galaxies out to this redshift. At the highest redshifts, the disc scalelength for the brightest galaxies with absolute magnitude between $-24$ and $-22$ is up to a factor 8 smaller compared to that in the local Universe. This observed scalelength decrease is significantly greater than the value predicted by a cosmological picture in which baryonic disc scalelength scales with the virial radius of the dark matter halo.

\end{abstract}

\begin{keywords}

galaxies: structure --
galaxies: evolution --
galaxies: high redshift
\end{keywords}

\section{INTRODUCTION}\label{intro}
A versatile observable used to establish the formation history of a galaxy disc is the exponential scalelength \citep{freeman70,lin87}. It relates directly to the disc mass and dark matter angular momentum, and thereby it can be used to derive the physical conditions of the gravitational collapse phase and galaxy merger remnants \citep[e.g., ][]{dalcanton97,dutton09}. It also dictates the evolution of substructures that may change the morphology of their host discs, caused by gravitational instabilities such as bulges, bars, and spiral arms \citep{bournaud07}. Disc scalelengths can also be converted into disc sizes and masses with the advantage that scalelength derivation is relatively independent on observed luminosity since it does not only rely on the detectability limit of a given surface brightness contour, e.g., effective radius of a galaxy, but it also relies on the brighter surface brightness contours on the face of the disc.

Galaxy sizes scale with redshift $z$ following three possible scaling relations: $H^{-1}$ at fixed halo circular velocity, $H^{-2/3}$ at fixed halo mass, and no evolution, where $H\equiv H(z)$ is the Hubble parameter which scales approximately as $(1+z)^{3/2}$ \citep{carroll92}. Observations below redshift $\sim 1.5$ suggest no disc size evolution out to these redshifts \citep[e.g., see][]{simard99,riv04,barden05}, whereas some studies have applied direct size measurement methods out to redshift 5 and beyond, and found support for disc evolution in agreement with the $H^{-2/3}$ model \citep{bouwens04,oesch10}. \citet{ferguson04} and \citet{hathi08}, on the other hand, argued for disc size evolution following the $H^{-1}$ model, which implies a constant galactic rotation curve in a cosmological picture in which baryonic disc scalelength scales with the virial radius of the dark matter halo \citep[e.g.,][]{mao98}. While these studies have investigated high redshift surveys carried out with the Hubble Space Telescope (HST), their results on disc size evolution are based on redshifts based on photometric measurements or drop-out techniques combined with direct size derivation methods. Direct size measurements are strictly limited by the detectability of the large lower surface brightness galaxies in any given magnitude in the sense that they are not able to pick out these galaxies and their correct sizes in the HST images.

To date, high redshift studies of scalelengths have found a factor two decrease at the mean redshift $\sim 2.5$ \citep{elmegreen05a} and a more detailed test bed for cosmological simulations of galaxy formation is still missing. Here, we present scalelengths for 686 galaxies out to redshift 5.8 based on archival data from major high redshift surveys carried out with the HST.

\begin{figure}
\centering\includegraphics[width=0.49\textwidth]{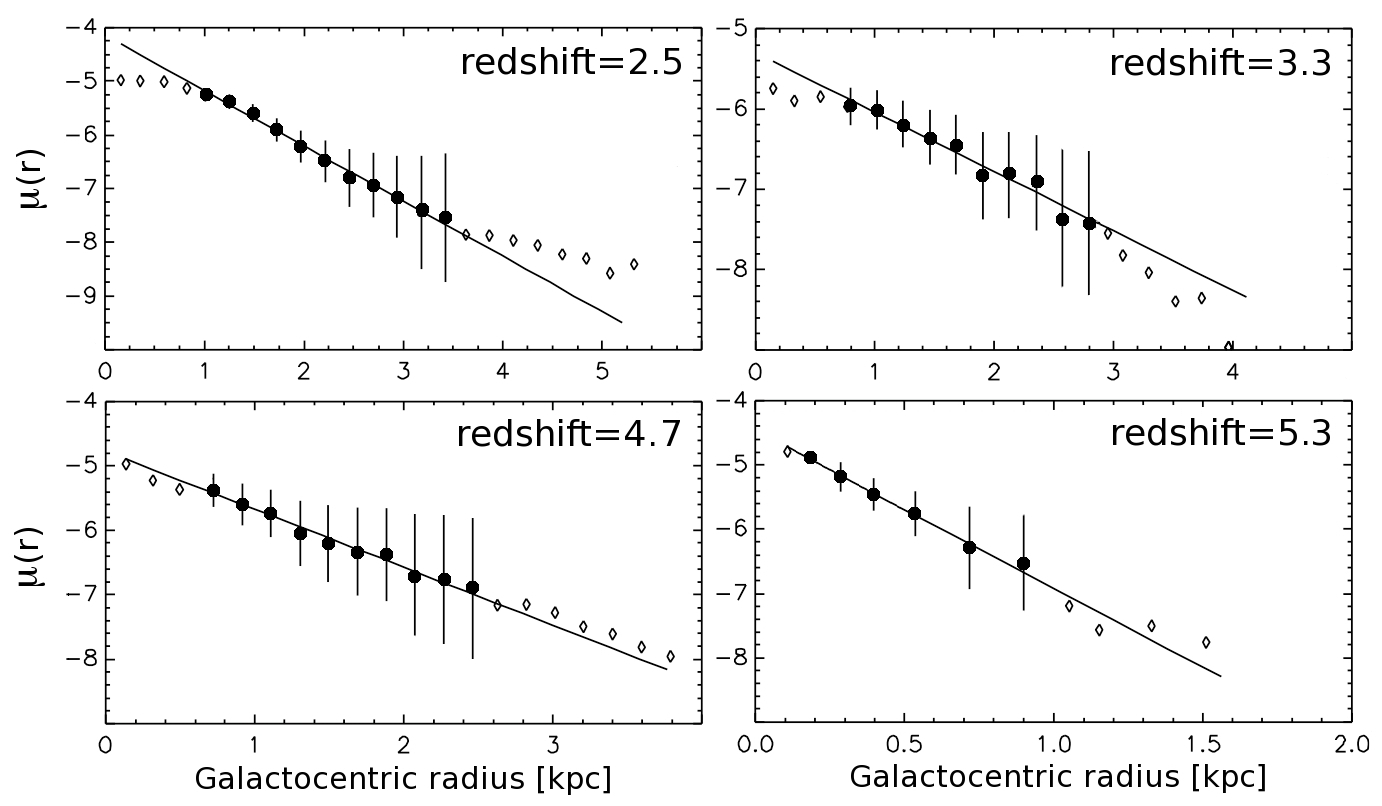}
\caption{Example light profiles from the HST sample with the best fit exponential profile (linear fit in log-scale). Black points represent the radii that were considered in each fit, and for clarity, the error bars are not displayed for all other points. Similar plots for the SDSS sample can be found in Fathi et al. (2010). The $y$-axis is given in arbitrary units.}
\label{fig:0}
\end{figure}

\section{DATA ANALYSIS}\label{sec:analysis}

We have analyzed all publicly available archival imaging data from the three deepest ever Hubble Space Telescope -- Advanced Camera for Surveys (HST-ACS) imaging surveys, namely the Great Observatories Origins Deep Surveys (GOODS) South and North \citep{giavalisco04} and the Hubble Ultra Deep Field (HUDF) \citep{beckwith06}. Our aim is to compare scalelengths for high redshift galaxies observed with the HST with our previous results for $\sim 30000$ $r-$band selected Sloan Digital Sky Survey \citep[SDSS; ][]{york00} galaxies out to redshift 0.3 \citep{fathi10a, fathi10}. The disc structure in the SDSS sample was quantified using our own customised one-dimensional light profile fitting, that was extensively tested and shown to deliver disc scalelengths very similar to previously published scalelengths derived by two-dimensional image fitting and multiple component fitting algorithms (further details can be found in Fathi et al. 2010 and Fathi 2010).

Images are available in the ACS F435W, F606W, F775W and F850LP filters for all three high redshift HST surveys. We used the best quality images in the F850LP filter ($z-$band) with Signal-to-Noise ratio $>$ 15 and select all galaxies (1642 in total) for which spectroscopic redshifts available in the literature \citep{barger99,cohen00,cowie04,vanzella04,vanzella06,vanzella08}. After selecting galaxies for which reliable images and solid or reliable redshift measurements (denoted by quality flags A or B in the online catalogues) are available, we investigated the presence of a dominant disc-like component in all galaxies. In the absence of a morphological classification catalogue for these samples, we used the diagnostic S\'{e}rsic profile fitting to the luminosity profile of each galaxy and derived a mean S\'{e}rsic index 1.3$\pm$0.4, in full agreement with high redshift observations \citep{toft05,vanderwel11}  and cosmological galaxy formation models \citep{mo98,dutton09}. The average S\'{e}rsic index of 1.3$\pm$0.4 strongly suggests the presence of a dominant exponential luminosity profile in our sample galaxies, often associated with a galactic disc. At all redshifts, approximately 25\% have S\'{e}rsic indices above 2, and less than 10\% above 4, thus, the contamination by E/S0 galaxies in our sample does not have a significant effect on the derived scalelengths.

We then quantified the structure of these exponential discs using the routines extensively tested in \citet{fathi10a} and  \citet{fathi10}. A brief summary follows: 

First an image tile is created centred on each galaxy, followed by the determination of the galaxy's isophotal major- and minor-axis and position angle, using SExtractor \citep{bertin96}. Stars and nuisance sources are masked out, and asymmetry measure and inclination are calculated. These can help identify and discard unreliable results and create a sample of high redshift galaxies with identical characteristics directly comparable to the SDSS. Each galaxy is divided into concentric rings with constant axis ratio as found in the previous step. A galactocentric light profile is created with the mean luminosity of each ring at every radius, fitted by $I(r)=I_0 e^{(r/r_d)}$, where $I(r)$ is the measured intensity at galactocentric each radius $r$, $I_0$ its value at the centre of the galaxy and $r_d$ the characteristic disc scalelength. The inner 50\% of each galaxy and the regions outside two galactic radii, are omitted from the fits, hence our fits are not affected by the presence of central bulges, circumnuclear starbursts, noisy or sky-dominated outskirts of galaxies. For all galaxies the choice of the outer fitting radius is justified by the small relative standard deviation for the derived average magnitudes at each radius, i.e., always less than one magnitude \citep{fathi10a}. Example fits are shown in Fig.~\ref{fig:0}.

\begin{figure*}
\centering\includegraphics[width=0.75\textwidth]{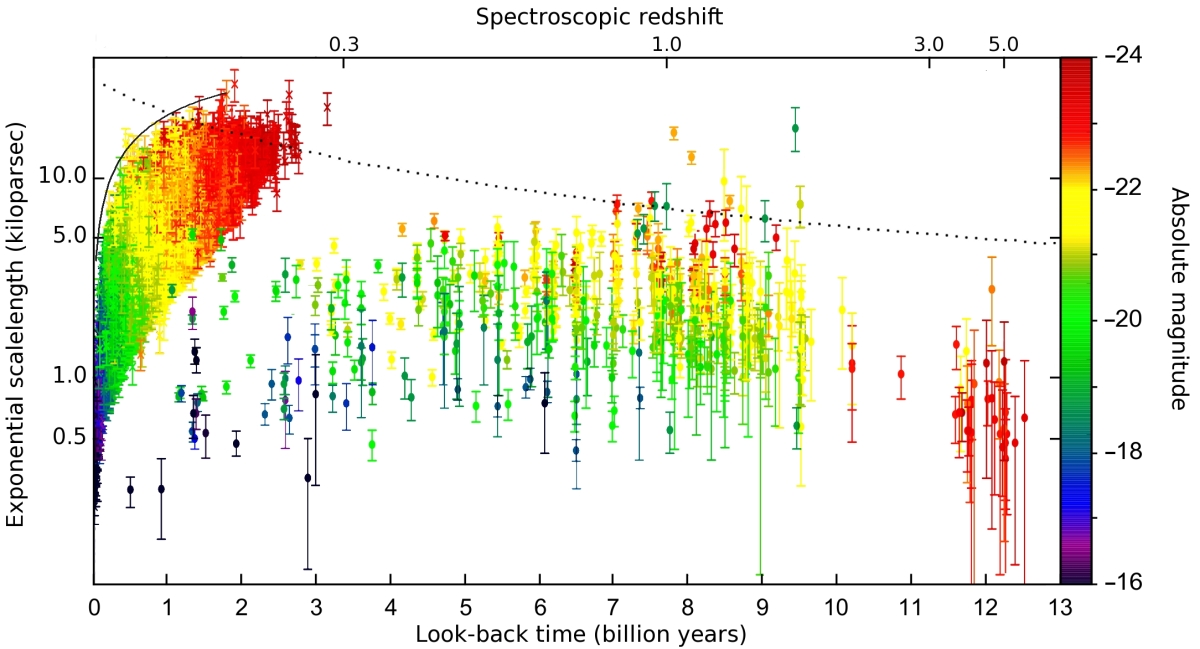}
\caption{Exponential disc scalelengths versus spectroscopic redshift over the last 12.5 Gyr. The arc of data points at the left shows the 30000 SDSS galaxies while the filled circles are the 686 HST galaxies at redshift $< 5.8$. The error bars illustrates the approximate scalelength uncertainties (35\%) due to observational effects. The dotted curve that illustrates the upper envelope for the decrease predicted for the bright nearby galaxies with large scalelengths. Out to redshift $\sim 1$--2, the scalelength decrease is in accordance with the theoretical predictions, whereas for the galaxies with absolute magnitudes between $-24$ and $-22$ and at redshifts $>2$, scalelengths decrease significantly greater than the prediction by the theoretical $H^{-2/3}$ model. The trend remains unchanged when using the photometric redshift sample with 1472 galaxies. The SDSS sample selection following fixed stamp size around each galaxy is illustrated by the solid curve at the top left of this diagram \citep[][]{fathi10a}.} 
\label{fig:1}
\end{figure*}

We discard galaxies for which 
$i)$ the inclination is greater than 60 degrees to avoid discs that have been diluted due to projection \citep{fathi10a}; 
$ii)$  asymmetry measure is greater than 0.35, to remove disturbed and unrelaxed systems \citep{shade95}; 
$iii)$ the disc light profile is represented by $\le 5$ data points; 
$iv)$ the least square fit has a relative error greater than unity.
Applying these criteria to the initial sample of 1642 galaxies with spectroscopic redshifts, we isolated a sub-sample of 686 galaxies with these properties identical to the 30000 SDSS galaxies studied in \citep{fathi10a}. The final HST sample spans out to redshift 5.8, corresponding to $\sim 12.5$ Gyr look-back time in a standard $\Lambda$CDM cosmology where $H_0=70$ km/s/Mpc, $\Omega_\Lambda =0.7$, and $\Omega_M=0.3$ (see Fig.~\ref{fig:1}).

\begin{figure}
\centering\includegraphics[width=0.40\textwidth]{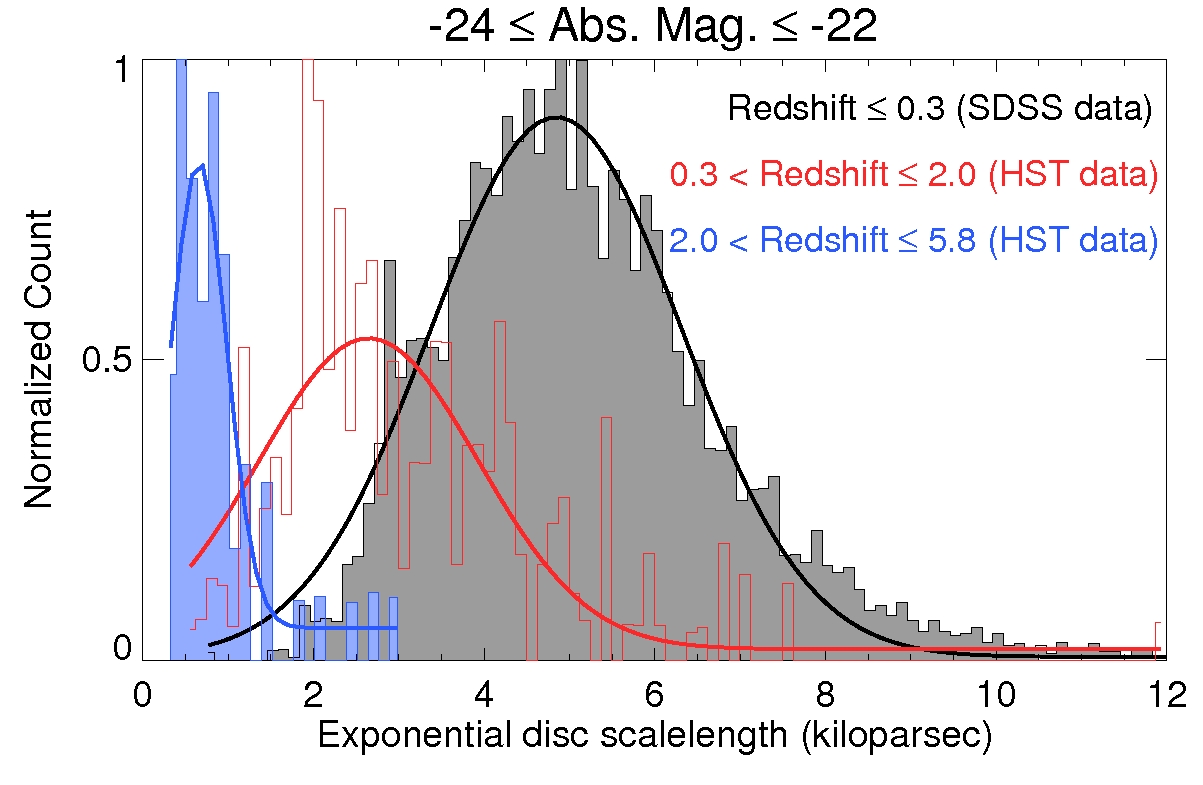}
\caption{Normalized distribution of scalelengths three absolute magnitude and redshift bins. The brightest absolute magnitude bin corresponding to the orange and red data points in Fig.~\ref{fig:1} are shown here. The SDSS data are illustrated by the black histogram (10040 galaxies) and HST data in red and blue histograms with 125 and 26 galaxies, respectively. Similar analysis performed on the photometric redshift galaxies yields the mean scalelengths presented in Table~1.}
\label{fig:2}
\end{figure}

We combine the scalelengths derived from the SDSS sample below redshift 0.3 with the HST sample, and divide the full set in three redshift bins (0--0.3; 0.3--2.0; 2.0--5.8). The nearby redshift bin includes 30210 galaxies with absolute magnitudes in the range ($-24,-14$). The intermediate bin includes 450 galaxies with absolute magnitudes between $-24$ and $-18$, and the highest redshift bin includes 26 galaxies all brighter than absolute magnitudes $-22$.  We further divide each bin into absolute magnitude subsets of two orders of magnitude, i.e., roughly corresponding to one order of magnitude in total galaxy mass. Each subset is thus populated by at least 26 galaxies. For each redshift and absolute magnitude bin, the mean galaxy disc scalelength values are quantified fitting a Gaussian function to the $V/V_{max}$ volume--corrected distribution histogram of each bin \citep[e.g., ][]{vanderkruit87}. Figure~\ref{fig:2} illustrates three such histograms with corresponding fits for the absolute magnitude bin ($-24,-22$).

We find that while brighter discs are typically more extended at any given redshift, the disc scalelength is decreasing at high redshift for any given intrinsic luminosity bin (see Fig.~\ref{fig:1}). At redshift $< 0.3$, we derive mean scalelength of $4.8 \pm 1.5$ kpc for galaxies with observed absolute magnitudes between $-24$ and $-22$. Their counterparts in the redshift range between 0.3 and 2.0 have a mean scalelength of $1.9 \pm 0.6$ kpc, and at redshifts between 2.0 and 5.8, $0.6 \pm 0.2$ kpc (see Fig.~\ref{fig:2}). Similar comparison for fainter intrinsic luminosity bins does not yield equally robust results due to the absence of spectroscopically confirmed galaxies with absolute magnitude fainter than $-22$ in the redshift bin 2.0--5.8. To examine this effect for fainter galaxies, we make use of all galaxies for which photometric redshifts are available in the literature \citep{coe06}, and present the mean scalelengths for each magnitude and redshift bin in Table~1 (see also Sec. \ref{sec:biases}).

\section{OBSERVATIONAL EFFECTS}\label{sec:effects}

\subsection{Biases}\label{sec:biases}

At the highest redshifts, the light observed in the $z-$band will have been emitted as ultraviolet photons which are produced by hotter stars. Ideally, a commensurable comparison of all scalelengths would have to account for wavelength shift due to a galaxy's redshift. Serious problems arise however, since at each redshift, we would be required to have detailed knowledge about dust properties, stellar populations, and galaxy evolution models for each morphological type. Such models are still highly uncertain at low redshift, and unavailable for high redshift galaxies. Studies of galaxies below redshift 1.0 have found that for a typical 10$^{11}$ M$_{\odot}$ galaxy, the salelength is $\sim 20$\% larger in the $B-$band as compared to the $K-$band \citep{peletier94, giovanelli02}. Thereof, wavelength dependencies cannot account for more than about 20\% of scalelength differences found here. Moreover, increased dust content makes a host galaxy more opaque which leads to an artificially larger bluer band scalelength when compared to red and/or infrared images \citep{valentijn90}. For a subsample of our 686 galaxies, for which reliable HST-ACS images with the F435W filter ($B-$band) are available, we have derived the scalelengths at both wavelength ranges. We found that at redshifts between 2.0 and 5.8 the $z-$band scalelengths are smaller than those measured in $B-$band by $\sim 20\%$. This indicates that the combined effects of stellar populations and dust add up quadraticly. 

At the redshift range between 2.0 and 5.8 the mean disc scaelength is $0.6 \pm 0.2$ kpc, corresponding to 2--3 HST-ACS pixels. Although the scalelength is not an actual galaxy size, it is imperative to quantify the effect of smearing on our derived scalelengths at high redshift. To ensure that the measured scalelengths do not contain any systematic redshift related bias due to differences in resolution for each redshift slice, we have artificially redshifted a sub-sample of 23 low-redshift galaxies, and re-derived their scalelengths at each epoch. All test galaxies were located at redshifts below 0.5, and we simulated their appearance at redshifts 2, 3, 4, and 5 using the redshift simulation code developed by \cite{epinat10} that simulates the reduction of resolution and rescaling of the point spread function of each image at the desired redshift. We found that, at the highest redshift, poorly resolved galaxies exhibit artificially lower scalelengths by at most 15\% of the initially observed low-redshift values, and S\'{e}rsic indices are artificially lowered by $\sim 20$\%, whereby the presence of an exponential component is re-confirmed.

\begin{figure}
\centering\includegraphics[width=0.49\textwidth]{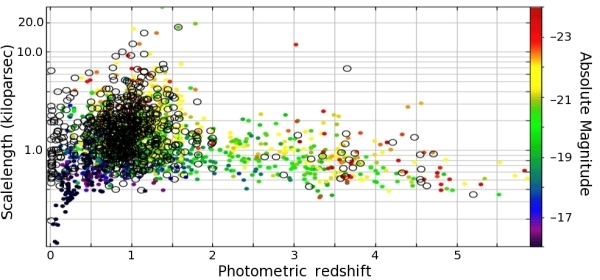}
\caption{Scalelength versus photometric redshift for the HUDF (coloured points) and GOODS galaxies (open circles). All GOODS galaxies above redshift 2.0 have absolute magnitudes below $-22$.}
\label{fig:photz}
\end{figure}

We further test the effect of cosmic variance, and find the same redshift-scalelength relation for each individual survey used here. We test the effect of Malmquist bias and include galaxies for which photometric redshifts are available. These galaxies are typically fainter, however their redshift measurements are not as reliable as we would require for our analysis, nevertheless, they still give us useful clues about a possible Malmquist bias effect. Deriving scalelengths and applying our selection criteria to a larger sample of 1472 galaxies with photometric redshifts from the HUDF \citep{coe06}, we find similar trends based on the distributions of the photometric redshift galaxies (see Table~1). We further note that the scalelength decrease in the GOODS galaxies \citep{bundy09} agrees with that found for the HUDF sample (see Fig.~\ref{fig:photz}), however, as we opt to use a homogenous set of photometric redshift data set and as the HUDF contains more galaxies at redshift $>2.0$ (for which we have reliable scalelengths), the values in Table~1 are based on the HUDF galaxies only.

In the local Universe, galaxies with absolute magnitude between $-18$ and $-20$ exhibit scalelengths a factor four smaller than their brightest counterparts. In the redshift range 2.0--5.8 galaxies in all these luminosity ranges have scalelengths that are significantly smaller than their local counterparts. In the intermediate redshift range, the Universe is populated by galaxies with intermediate scalelength values (see Table~1). The photometric redshift sample further shows that scalelengths remain redshift-invariant for systems fainter than absolute magnitude $> -18$ and below redshift 1.5.

High redshift galaxy population may differ significantly from galaxies in the local Universe. The effects of stellar population and dust content on the measured scalelengths are of the order of 20\%. The effect of spatial resolution and sampling accounts for an additional 15\%. Adding all these effects quadraticly, the resulting scalelengths at high redshift are $\sim 35\%$ lower than their intrinsic values.

\subsection{Surface Brightness Limitation}\label{sec:sb}

Due to the degeneracy between surface brightness selection effect and direct luminosity and size measurements, the HST images analyzed here would not allow direct detection of galaxies with half-light radii above $\sim 25$ pixels at $z-$band AB magnitude fainter than 24, and above $\sim 15$ pixels at $z-$band magnitudes fainter than 26. As found by, e.g., \citet{bouwens06}, significant loss of the large low surface brightness galaxies are expected to affect the data at $z-$band AB magnitudes fainter than $\sim 27$ and axial ratio analysis of disc galaxies out to redshift 4 has revealed that only 25\% of the spirals larger than 10 pixels are lost below the HST detection limit \citep{elmegreen05b}. These limits are significantly fainter than the $z-$band AB magnitude limit of our spectroscopic and photometric redshift sample, $\sim 24$ and $\sim 26$, respectively.

\begin{table*}
 \label{tab}
 \centering\tabcolsep10pt
   \begin{threeparttable}
      \begin{tabular}{|c|c|c|c|}
     \hline
     \multirow{2}{*}{\textbf{Absolute Magnitude}} &
     Scalelength (kiloparsec) & Scalelength (kiloparsec)&  Scalelength (kiloparsec)\\
         & $0<$ redshift $<0.3$ & $0.3<$ redshift $<2.0$ & $2.0<$ redshift $<5.8$ \\ \hline\hline
 ($-24, -22$)         & $4.8 \pm 1.5$ \colorbox{mycol}{\parbox{2.25cm}{\center \footnotesize 10040 galaxies}}
                             & $2.0 \pm 0.6$ \colorbox{mycol}{\parbox{2.25cm}{\center \footnotesize 146 galaxies}}
                             & $0.7 \pm 0.3$ \colorbox{mycol}{\parbox{2.25cm}{\center \footnotesize 56 galaxies}}\\ \hline
 ($-22, -20$)         & $2.7 \pm 1.2$ \colorbox{mycol}{\parbox{2.25cm}{\center \footnotesize 17802 galaxies}}
                             & $1.3 \pm 0.5$ \colorbox{mycol}{\parbox{2.25cm}{\center \footnotesize 514 galaxies}}
                             & $0.7 \pm 0.2$ \colorbox{mycol}{\parbox{2.25cm}{\center \footnotesize 110 galaxies}}\\ \hline
 ($-20, -18$)         & $1.2 \pm 0.4$ \colorbox{mycol}{\parbox{2.25cm}{\center \footnotesize 1886 galaxies}}
                             & $0.8 \pm 0.3$ \colorbox{mycol}{\parbox{2.25cm}{\center \footnotesize 281 galaxies}}
                             & $0.6 \pm 0.2$ \colorbox{mycol}{\parbox{2.25cm}{\center \footnotesize 60 galaxies}}\\ \hline
    \end{tabular}
\caption{We test the validity of our results based on spectroscopic redshifts by expanding the sample to include galaxies for which photometric redshifts are available. All values presented here are based on Gaussian fits to the $V/V_{max}$ volume--corrected scalelength distribution of the SDSS sample combined with the 1472 photometric redshift galaxies in the HUDF. In each cell, the number of galaxies are written at the right. The errors are the standard deviations of the Gaussian fits (c.f. Fig.~\ref{fig:2}). The weighted means remain within the uncertainties when using both HUDF and GOODS galaxies (c.f., Fig.~\ref{fig:photz}).}
\end{threeparttable}
\end{table*}

\section{DISCUSSION}\label{sec:discussion}

Our results agree with previous works postulating that the first generation galaxies are expected to be gas-rich, and when they undergo mergers, they leave behind substantially rotating remnants and thereby galaxy discs are formed early in the Universe \citep{wuyts10}. The discs then continue to grow to transform into the Hubble sequence observed in the local Universe. Comparing our derived scalelengths with those measured by means of direct size measurements, we find that direct size measurements typically deliver over-estimated values due to their strong dependancy on the observed luminosity in the sense that they do not pick out the larger lower surface brightness galaxies \citep[e.g.,][]{bouwens04}. Further advantage of deriving scalelengths as opposed to direct size measurements for high redshift galaxies can be seen in the comparison with predictions from cosmological galaxy formation simulations with the computational resolution that allows the calculation of disc scalelengths \citep[e.g., ][]{ceverino10, dutton09}. On the observational side, to date, scalelength measurements have been usually based on inhomogeneous data sets from the local Universe. It is only two years ago that scalelengths were derived for a comprehensive sample of 30000 SDSS galaxies out to redshift 0.3 \citep{fathi10a}, while studies beyond redshift $\sim 1.5$ were limited to a handful of objects \citep{vanderkruit11}. Here we have compared our results from the SDSS sample with scalelengths derived from a sample of 686 galaxies with spectroscopic redshifts $\leq 5.8$. While we use 1472 photometric redshift galaxies to test against Malmquist bias, our results are based on solely spectroscopic redshift measurements, whereas all previous works dealing with size evolution of disc galaxies, have been based on photometric redshifts or drop-out techniques. 

Our results confirm the presence of a dominating exponential components in galaxies out to redshift 5.8 \citep[c.f., ][]{riv06}, and we find that at redshifts between 2.0 and 5.8 (corresponding to approximately 10--12.5 Gyr ago), bright disc galaxies had scalelengths up to a factor 8 smaller than their local counterparts. A similar trend for galaxies fainter than absolute magnitude $-22$ cannot be confirmed. The observed decrease out to redshift 5.8 is significantly greater than the value predicted by the cosmological picture in which baryonic disc scalelength scales with the virial radius of the dark matter halo. At the highest redshifts, direct size measurement methods suggest a rather slow disc size evolution, whereas our results show a disc size growth speedier than that predicted by the $H^{-2/3}$ model \citep[e.g.,][]{bouwens04}, and slower than that predicted by the $H^{-1}$ model \citep[e.g.,][]{ferguson04}. Our findings suggest that the bulk of galaxy growth occurred in the first 1--3 Gyr after Big Bang, i.e., right after the formation of the first galaxies, and that a combination of both models should be considered in cosmological galaxy formation models.

The observational picture presented here offers a uniquely quantified and newly examined scenario for modellers of galaxy formation and evolution within a cosmological paradigm. Applying a similar analysis to wide area high redshift surveys such as CANDELS \citep{grogin11} will help building larger sample and improve measurement of galaxy structural parameters at high redshifts.

\section*{ACKNOWLEDGMENTS}
We thank the anonymous referee for particularly constructive comments and insightful suggestions and guidance. All data used in this work are public and made available by the GOODS and HUDF teams. M.G. thanks the European Southern Observatory headquarters in Garching for their hospitality during a visit when parts of this work was carried out. K.F. acknowledges support from the Swedish Research Council (Vetenskapsr\aa det). This work made use of TOPCAT written by Mark B. Taylor.

\label{lastpage}

\end{document}